\documentclass[journal=jpcl,manuscript=article,layout=twocolumn]{achemso}

\usepackage[version=3]{mhchem} 
\usepackage{multirow}
\usepackage{amsmath}
\usepackage{amssymb}
\usepackage{graphicx}
\usepackage{soul} 
\usepackage[normalem]{ulem}

\usepackage[dvipsnames]{xcolor}

\author{Marc Rico-Pasto}
\affiliation[]
{ Small Biosystems Lab, Condensed Matter Physics Department, University of Barcelona, C/Mart\'{\i} i Franqu\'es 1, 08028 Barcelona, Spain}

\author{Anna Alemany}
\affiliation[]
{Department of Anatomy and Embryology, Leiden University Medical Center, 2300 Leiden, The Netherlands}

\author{Felix Ritort}
\email{ritort@ub.edu}
\affiliation[]
{ Small Biosystems Lab, Condensed Matter Physics Department, University of Barcelona, C/Mart\'{\i} i Franqu\'es 1, 08028 Barcelona, Spain}

\title {Force-Dependent Folding Kinetics of Single Molecules with Multiple Intermediates and Pathways}

\let\oldmaketitle\maketitle
\let\maketitle\relax

\begin{document}
\maketitle

\twocolumn[
\begin{@twocolumnfalse}
\oldmaketitle
\begin{abstract}
Most single-molecule studies derive the kinetic rates of native, intermediate, and unfolded states from equilibrium hopping experiments. Here, we apply Kramers kinetic diffusive model to derive the force-dependent kinetic rates of intermediate states from non-equilibrium pulling experiments. From the kinetic rates, we also extract the force-dependent kinetic barriers and the equilibrium folding energies. We apply our method to DNA hairpins with multiple folding pathways and intermediates.  The experimental results agree with theoretical predictions. Furthermore, the proposed non-equilibrium single-molecule approach permits us to characterize kinetic and thermodynamic properties of native, unfolded, and intermediate states that cannot be derived from equilibrium hopping experiments. 
\end{abstract}
\end{@twocolumnfalse}
]

\begin{figure*}[ht]
    \centering
    \includegraphics{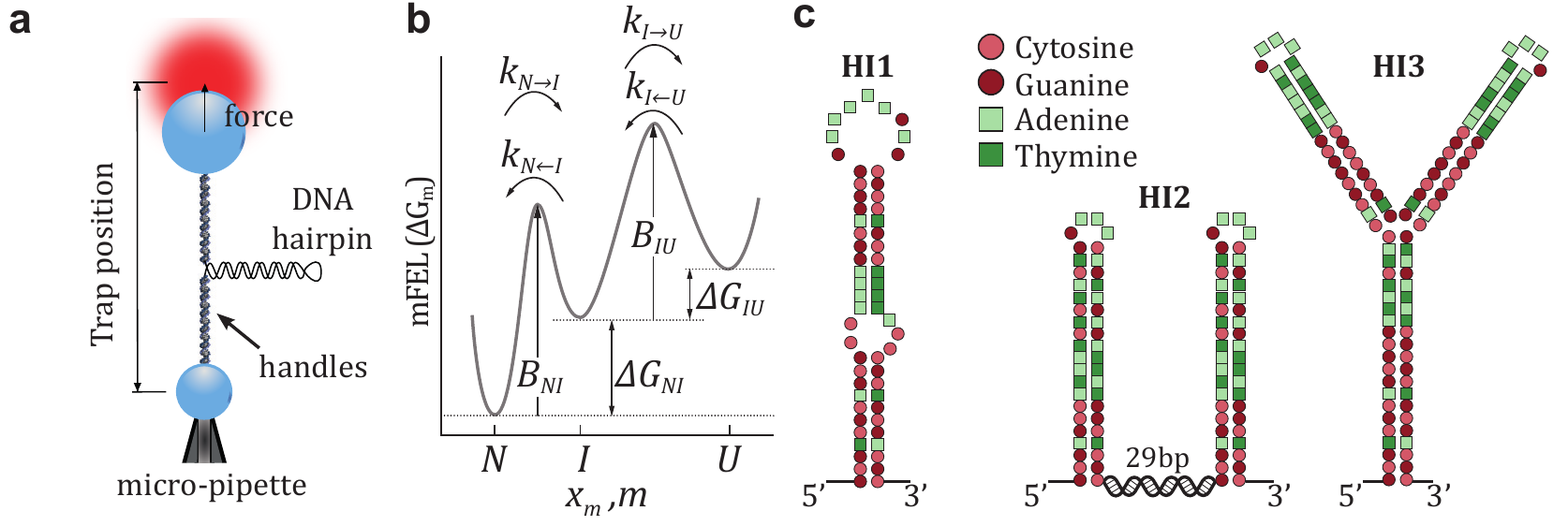}
    \caption{\textbf{Experimental setup and DNA sequences} (a) Schematics of a pulling experiment with optical tweezers. The DNA hairpin is tethered between two beads using double-stranded DNA handles. One bead is fixed by air suction on the tip of a micro-pipette while the other is controlled by the optical trap. (b) Illustrative mFEL with an intermediate state. The unfolding and folding  kinetic rates and barriers are indicated. (c) Sequences of the studied DNA molecules.}
    \label{fig1}
\end{figure*}

Some nucleic acids and proteins require intermediate or partially folded configurations to perform their biological function. For example, RNA riboswitches are regulatory molecules that induce or repress gene transcription depending on their conformation \cite{serganov2013}; RNA thermometers act like lockers whose ribosomal binding site becomes accessible only at high temperatures when they partially unfold \cite{kortmann2012, krajewski2014}; and proteins fold into the native structure by forming intermediate folding units (foldons) \cite{Maity2005, Baldwin2017}. Therefore, a quantitative characterization of the dynamical formation of intermediates is a critical step towards the elucidation of many molecular processes. Accordingly, it is of high interest to develop accurate tools to investigate the thermodynamics and kinetics of partially folded domains occurring in biomolecules.

Single-molecule methods provide an ideal ground to experimentally address these questions since they allow us to sample transient molecular states with high temporal ($\sim$ ms) and spatial resolution ($\sim$ nm) \cite{miller2017}. In particular, atomic force microscopy \cite{ruggeri2016, sluysmans2021}, magnetic \cite{kriegel2017, dahal2020} and optical tweezers \cite{huguet2017, choudhary2019, bustamante2021}, permit us to pull on individual molecules and to monitor unfolding/folding reactions from the recorded changes in extension, the reaction coordinate in these experiments\cite{best2005, best2008}. 

Single-molecule techniques have been used to characterize intermediates in a wide variety of molecular systems, from protein folding \cite{wang2015, yu2017, aviram2018} and binding metal-metalloproteins \cite{zheng2013, zheng2014, goldman2015} to RNA and DNA folding \cite{hyeon2006, hyeon2007, chen2007}, G-quadruplex DNA formation \cite{long2013, yu2013}, DNA duplexes formation with base-pair mismatches \cite{yang2018}, and synthetic molecular foldamers and shuttles \cite{sluysmans2018, naranjo2018}. Moreover, upon misfolding, molecular intermediates have also been shown to play a role, e.g., in neuronal calcium sensors \cite{heidarsson2014}.

Dynamic force spectroscopy studies are often performed in equilibrium conditions, e.g., in hopping experiments \cite{Wen2007, Manosas2007, gebhardt2010, elms2012, neupane2016, bustamante2020}. There, the control parameter (e.g., trap position in optical tweezers, Fig.~\ref{fig1}a) is kept fixed as the molecule executes thermally-driven transitions between different molecular states. In such experiments, the unfolding and folding kinetics are derived from the average lifetime of each state \cite{Forns2011, rico2018}. However, equilibrium experiments are strongly limited by the height of the kinetic barrier, $B$, mediating transitions between contiguous states along the molecular free energy landscape (mFEL) (Fig.~\ref{fig1}b). A too high kinetic barrier ($B \gg k_BT$, being $k_B$ the Boltzmann constant and $T$ the temperature) prevents molecular transitions over measurable timescales, leading to inefficient sampling of the conformational space. Instead, non-equilibrium experiments facilitate transitions over large kinetic barriers, providing an alternative and efficient way to sample the mFEL. Examples are jump experiments where a system is driven to a new state by suddenly changing an external parameter (such as temperature, force, pH, etc..) and system's relaxation monitored \cite{Li2006, Causgrove2006, Wirth2015}.

Two widely used phenomenological approaches to extract equilibrium information from pulling experiments are the Bell-Evans (BE) \cite{bell1978, evans1997, evans2001} and the kinetic diffusive (KD) models. The BE model describes mechanically induced folding/unfolding transitions as thermally activated processes over a transition state energy barrier. The BE model assumes that, for a fixed transition state position, the height of the kinetic barrier decreases linearly with the applied force, $B=B_0-fx^{\dagger}$ (being $x^{\dagger}$ the distance from the departure state to the transition state). This assumption is relaxed in the KD model, which considers the folding reaction as a diffusive process in a one-dimensional force-dependent mFEL (Fig.~\ref{fig1}b). While the BE model only considers the height and position of the transition state, the full description of the mFEL in the KD model requires the knowledge of all the partially folded intermediate conformations. The advantage of the KD is the high predictive power. The same experimental data can be readily employed to extract additional information about the mFEL without the need to adopt the assumptions of the BE model. The KD model has been applied to study the folding kinetics of two-state nucleic acid hairpins and proteins \cite{gebhardt2010, gupta2011, neupane2015, neupane2016}.

A useful method based on the KD model is the Continuous Effective Barrier Approach (CEBA). Originally introduced to study RNA hairpins \cite{manosas2006}, it has been later applied to extract the elastic properties of short RNA hairpins at different ionic conditions \cite{bizarro2012}, the thermodynamic and kinetic properties of protein Barnase \cite{alemany2016}, and DNA hairpins with different mechanical fragilities \cite{alemany2017}. In CEBA, the force-dependent effective barrier between the native ($N$) and the unfolded state ($U$), $B_{NU}(f)$, is derived by imposing detailed balance between the unfolding ($k_{N\to U}(f)$) and folding ($k_{N\leftarrow U}(f)$) kinetic rates:
\begin{subequations}
\begin{equation} \label{k_Unf}   
k_{N\to U}(f)=k_0 \exp{\Bigl(-\frac{B_{NU}(f)}{k_BT}\Bigr)}
\end{equation}
\begin{equation} \label{k_Ref}   
k_{N\leftarrow U}(f)=k_{N\to U}(f) \exp{\Bigl(\frac{\Delta G_{NU}(f)}{k_BT}\Bigr)}
 \ \ .
\end{equation}
\end{subequations}
Here $k_0$ is the attempt rate and $\Delta G_{NU}(f)$ is the folding free energy at force $f$,
\begin{equation}
  \Delta G_{NU}(f)=\Delta G^0_{NU}-\int_0^f (x_U(f') - x_N(f'))df' 
   \label{G0_int}
\end{equation}
with $\Delta G^0_{NU}$ the folding free energy difference between $N$ and $U$ at zero force, and $-\int_0^f x_{U(N)}(f')df'$ the free energy decrease upon stretching the molecule in state $U(N)$ at force $f$. The elastic response of $U$ and $N$ are modeled using the Worm-Like Chain and Freely-Jointed Chain models \cite{ViaderGodoy2021}. Eqs.(\ref{k_Unf}, \ref{k_Ref}) are conveniently rewritten as, 
\begin{subequations}
\begin{equation} \label{Beff_Unf}   
\frac{B_{NU}(f)}{k_BT} = \log k_0 -\log{k_{N\to U}(f)}
\end{equation}
\begin{equation} \label{Beff_Ref}   
\frac{B_{NU}(f)}{k_BT} = \log k_0 -\log{k_{N\leftarrow U}(f)}  + \frac{\Delta G_{NU}(f)}{k_BT}
\end{equation}
\end{subequations}
Therefore, by equating Eq.(\ref{Beff_Unf}) and Eq.(\ref{Beff_Ref}), the difference between $-\log k_{N\to U}(f)$ and $-\log k_{N\leftarrow U}(f)-(1/k_BT)\int_0^f (x_U(f') - x_N(f'))df'$ equals $\Delta G^0_{NU}$ (c.f. Eq.(\ref{G0_int})). This permits us to derive the folding free energy $\Delta G^0_{NU}$ if the elastic response $(x_U(f) - x_N(f))$ is known. Moreover, we extract $k_0$ by comparing the experimental profile of $B_{NU}(f)-\log k_0$ with the theoretically predicted $B_{NU}(f)$ by the KD model \cite{alemany2016, alemany2017}. For a DNA hairpin, the latter is given by \cite{kramers1940,zwanzig2001,hyeon2007b} (a derivation can be found in section S1 of Supp. Info.):
\begin{equation}\label{Beff_TheoNU}
    \frac{B_{NU}(f)}{k_BT} =  \log \left(\sum_{m=0}^{M}\sum_{m'=0}^m e^{\left( \frac{\Delta G_m(f) - \Delta G_{m'}(f)}{k_BT}\right)} \right)
\end{equation}
where the double sum runs over all hairpin configurations, labeled by $m$ and $m'$, and $M$ being the total number of base pairs (bp).


\begin{figure*}
    \centering
    \includegraphics{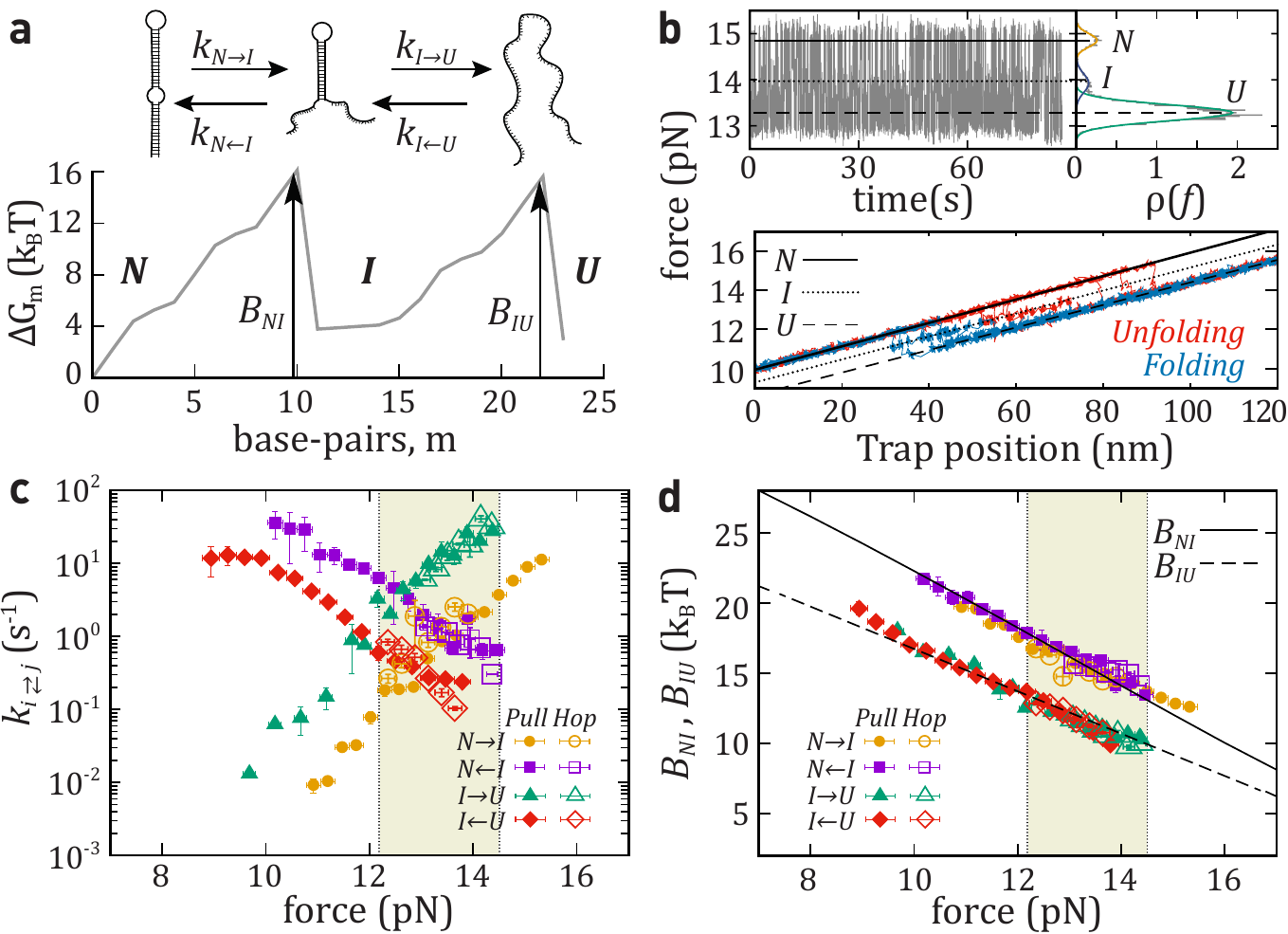}
    \caption{\textbf{Unfolding-folding kinetics of a single intermediate (hairpin HI1).} (a) Top: Schematic unfolding and folding pathway for HI1. Bottom: mFEL ($\Delta G_m$) as a function of the number of unfolded base-pairs ($m$) at $15$pN. The barriers, $B_{NI}$ and $B_{IU}$, are highlighted (black arrows). (b) Top-left: Force versus time trace measured for HI1. Top-right: Histogram of the force signal used to recognize the three states ($N$, $I$, and $U$). Bottom: Five unfolding (red) and folding (blue) FDCs (pulling speed equals $100$nm/s,  each trajectory taking $\sim$ 2s.). Force branches for states $N$ (black solid line), $U$ (black dashed line) and $I$ (dotted line). (c) Kinetic rates of unfolding: $k_{N \to I}$ (yellow circles) and $k_{I \to U}$ (green triangles); and folding: $k_{N \leftarrow I}$ (purple square) and $k_{I \leftarrow U}$ (red diamond). Kinetic rates derived from non-equilibrium (solid symbols) and equilibrium (empty symbols) experiments. (d) Barriers mediating transitions between $N$ and $I$ (black solid line) and between $I$ and $U$ (black dashed line) as predicted from Eqs.(\ref{Beff_TheoNI},\ref{Beff_TheoIU}) compared with the experimental results (symbols). Shaded regions in panels (c) and (d) show the range of forces where kinetic rates can be measured in equilibrium hopping experiments. Results are the average over four different molecules and the error bars correspond to the statistical errors.}
    \label{fig2}
\end{figure*}

CEBA has been mostly applied to molecules with two distinct molecular states, i.e., $N$ and $U$, separated by a kinetic barrier \cite{manosas2006, bizarro2012, alemany2016, alemany2017}. Here, we extend CEBA (hereafter referred to as eCEBA) to investigate molecular reactions involving intermediate kinetic states from non-equilibrium pulling experiments. We use optical tweezers to pull DNA hairpins with one, two and three intermediates (Fig.~\ref{fig1}a). The existing knowledge about DNA thermodynamics \cite{santalucia1998, zuker2003, huguet2010, huguet2017} allows us to accurately predict the force-dependent kinetic barriers of arbitrary sequences, facilitating the comparison between theory and experiments. The chosen examples cover situations often encountered in macro-molecular folding.

{\bf A single intermediate and folding pathway.} The first DNA hairpin (denoted as HI1) has an internal loop in the stem (Fig. \ref{fig1}c) that stabilizes an intermediate ($I$) upon folding/unfolding, as shown in the theoretical prediction of the mFEL\cite{mossa2009, huguet2010, santalucia1998, zuker2003} calculated at $15$pN  (Fig. \ref{fig2}a). Partial folding and unfolding connecting states $N$, $I$, and $U$ can be observed as sudden drops and rises of force, respectively, in hopping (equilibrium) and pulling (non-equilibrium) experiments (Fig. \ref{fig2}b). In hopping experiments, the molecule is held at a fixed trap position (distance), and each observed level of force corresponds to a different state (Fig. \ref{fig2}b, top). In pulling experiments, the trap position is moved back and forth at a constant speed and the molecule is repeatedly folded and unfolded. The different force branches observed in the force-distance curves (FDCs) arise from the elastic response of the hairpin in each state (black lines in Fig. \ref{fig2}b, bottom). Let us note that fast hopping events are missed in force feedback protocols with optical tweezers, underestimating the kinetic rates \cite{rico2018}. Then, a proper comparison between hopping and pulling should be done in the same experimental condition (either controlling force or distance).

In hopping experiments, the kinetic rates $k_{N\to I}(f)$, $k_{I\to U}(f)$, $k_{N\leftarrow I}(f)$, and $k_{I \leftarrow U}(f)$ are derived from the lifetime of each state (empty symbols in Fig. \ref{fig2}c). In pulling experiments, we determine them from the survival probabilities of each state along the unfolding and folding FDCs. The methodology to determine the survival probabilities for molecules with an arbitrary number of intermediates is very general. For the single intermediate case it is as follows. 
Firstly, we set a threshold force, $f_{th}$, and measure the first rupture (formation) event taking place at a force above (below) $f_{th}$ for each unfolding (folding) trajectory. Next, we classify the force events as: $f^i_\to$ and $f^i_{\leftarrow}$, where $i=N,~I$ or $U$ indicates the molecular state at $f_{th}$ and the arrow indicates the direction of the FDCs: unfolding ($\to$) or folding ($\leftarrow$). Note that for $i=I$, $f^I_\to$ and $f^I_\leftarrow$ comprise both rupture and formation events indistinguishably, while $f^N_\to$ and $f^N_\leftarrow$ only contain rupture events and $f^U_\to$ and $f^U_\leftarrow$ only contain formation events. From $f^i_\to,f^i_{\leftarrow}$, we calculate the force-dependent survival probabilities conditioned to $f_{th}$ along the unfolding ($P_\to^N(f|f_{th}),~P_\to^I(f|f_{th}), ~P_\to^U(f|f_{th})$) and folding ($P_\leftarrow^N(f|f_{th}),~P_\leftarrow^I(f|f_{th}), ~P_\leftarrow^U(f|f_{th})$) trajectories: 
\begin{subequations}
\begin{equation} \label{P_Unf}   
P_\to^{i}(f|f_{th}) = 1 - \frac{n(f_{th} < f_{\to}^i < f)}{n^i_{\to}}
\end{equation}
\begin{equation} \label{P_Ref}   
P_\leftarrow^{i}(f|f_{th}) = 1 - \frac{n(f_{th} > f_{\leftarrow}^i > f)}{n^i_{\leftarrow}}
\end{equation}
\end{subequations}
%
where $n(f_{th} < f_{\to}^i < f)$ ($n(f_{th} > f_{\leftarrow}^i > f)$) denotes the number of events during unfolding (folding) leaving state $i$ for the first time between $f_{th}$ and $f$, and $n^i_\to$ ($n^i_\leftarrow$) is the total number of trajectories with state $i$ observed at $f_{th}$. Note that by construction, $P_{\to(\leftarrow)}^{i}(f_{th}|f_{th})=1$. By repeating the analysis for different values of $f_{th}$, we reconstruct $P_\to^{i}(f|f_{th})$ and $P_\leftarrow^{i}(f|f_{th})$ for different values of $f_{th}$ and $f$. The survival probabilities satisfy the following master equations:  
\begin{subequations}
\begin{equation} \label{eq:kNI}   
  \frac{\partial P_{{\rightleftarrows}}^N(f|f_{th})}{\partial f} = {\mp} \frac{k_{N \to I}(f)}{r} P_{{\rightleftarrows}}^N(f|f_{th})
\end{equation}
\begin{equation} \label{eq:kI}   
\frac{\partial P_{{\rightleftarrows}}^I(f|f_{th})}{\partial f} = {\mp} \frac{k_{N \leftarrow I}(f) + k_{I \to U}(f) }{r} P_{{\rightleftarrows}}^I(f|f_{th}) 
\end{equation}
\begin{equation} \label{eq:kUI}   
\frac{\partial P_{{\rightleftarrows}}^U(f|f_{th})}{\partial f} = {\mp} \frac{k_{I \leftarrow U}(f)}{r} P_{{\rightleftarrows}}^U(f|f_{th})\, .
\end{equation}
\end{subequations}
%
where $k_{N \to I}(f)$, $k_{N \leftarrow I}(f)$ and $k_{I \leftarrow U}(f)$ are the kinetic rates between the states, and $r=|df/dt|$ is the constant loading rate. The -(+) sign in the rhs denote the unfolding (folding) processes. 

\begin{figure}
    \centering
    \includegraphics{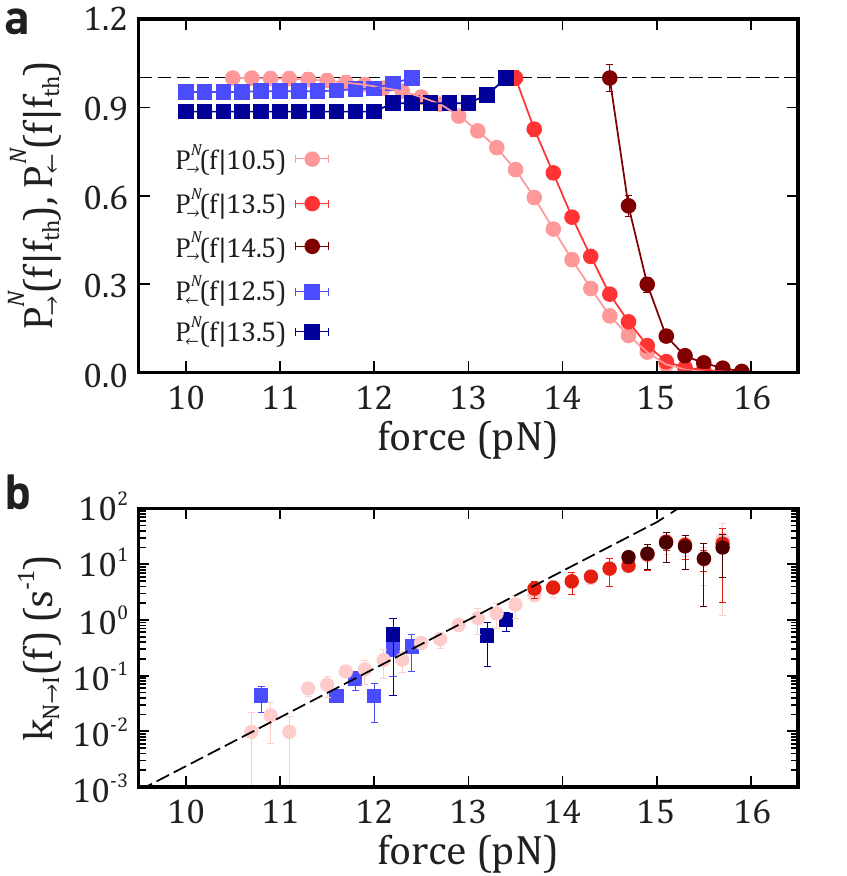}
    \caption{\textbf{Survival probability of \textit{N} and derived kinetic rate \textit{N $\to$ I}} (a) Survival probability of $N$ in the unfolding (red circles) and folding (blue squares) processes. (b) Derived $k_{N \to I}(f)$ using Eq.\eqref{eq:kNI}. Data fall on the same line. Error bars are the statistical errors over five molecules. The black dashed line is the prediction by the KD model.}
    \label{fig3}
\end{figure}

For a Markovian system, Eqs. (\ref{P_Unf}-\ref{eq:kUI}) give estimates of $k_{N \rightleftarrows I}$, $k_{I \rightleftarrows U}$ that are independent of the value $f_{th}$ and the process $\to(\leftarrow)$. Therefore, by merging results obtained at different $f_{th}$ and $\to(\leftarrow)$ we optimize the available data improving kinetic rates estimates. In Figure \ref{fig3}a, we show $P_{{\rightleftarrows}}^N(f|f_{th})$ for three values of $f_{th}$ for $\to$ and two $f_{th}$ values for $\leftarrow$ processes. The corresponding $k_{N \to I}(f)$ derived from Eq.\eqref{eq:kNI} are compatible with each other (Fig. \ref{fig3}b).

A similar procedure is used to determine $k_{I\leftarrow U}(f)$.  To decouple $k_{N\leftarrow I}(f)$ from $k_{I\to U}(f)$ in Eq.\eqref{eq:kI}, we use the following relation:
\begin{equation}
    \frac{k_{I \to U}(f)}{k_{N \leftarrow I}(f)} = \frac{\phi_{I \to U}(f)}{\phi_{N \leftarrow I}(f)}=\frac{\phi_{I \to U}(f)}{1-\phi_{I \to U}(f)}
\end{equation}
where $\phi_{I \to U}(f)$ and $\phi_{N \leftarrow I}(f)$ are the fraction of transitions leaving $I$ towards $U$ and $I$ towards $N$, respectively, at force $f$ ($\phi_{I \to U}(f)+\phi_{N \leftarrow I}(f)=1$). These fractions are experimentally measured on a force window $\Delta f = 0.1$pN.

Figure \ref{fig2}c shows a good agreement between kinetic rates recovered from hopping (empty symbols) and pulling experiments (solid symbols). Notably, the force range where transitions are observed in hopping (highlighted in yellow) is narrower compared to that from pulling experiments. This shows that non-equilibrium pulling experiments provide kinetic rates over a wider force range. 

Next, we use eCEBA to determine the effective barriers $B_{NU}(f)$ and $B_{IU}(f)$ from the kinetic rates by generalizing Eqs.(\ref{Beff_Unf}, \ref{Beff_Ref}) to states $N,I,U$:
\begin{subequations}
\begin{equation} \label{Beff_Unf_exp}   
\frac{B_{ij}(f)}{k_BT} = \log k_0^{ij} -\log{k_{i\to j}(f)}
\end{equation}
\begin{equation} \label{Beff_Ref_exp}   
\frac{B_{ij}(f)}{k_BT} = \log k_0^{ij} -\log{k_{i\leftarrow j}(f)}  + \frac{\Delta G_{ij}}{k_BT}\ \ .
\end{equation}
\end{subequations}
with $i,j\in\{N,I,U\}$, $k_{i\to j}$, $k_{i\leftarrow j}$ the unfolding and folding kinetic rates between $i$ and $j$, and $k_0^{ij}$ the attempt rate. $\Delta G_{ij}$ equals $\Delta G^0_{ij}-\int_0^f (x_j(f') - x_i(f'))df'$, where $\int_0^f x_{k}(f')df'$ is the energy cost to stretch state $k$ up to force $f$, and $\Delta G^0_{ij}$ the folding free energy difference at zero force between states $i$ and $j$. 

By imposing continuity between the two expressions for $B_{ij}(f)/k_BT-\log k_0^{ij}$ in Eqs.(\ref{Beff_Unf_exp}, \ref{Beff_Ref_exp}) we derive $\Delta G^0_{ij}$ for $ij=NI$ and $ij=IU$. We also estimate the attempt frequencies $k_0^{ij}$ by matching the experimental results for $B_{ij}(f)$ with the theoretical Kramers prediction, calculated as \cite{kramers1940}: 
\begin{subequations}
\begin{equation}\label{Beff_TheoNI}
    \frac{B_{NI}(f)}{k_BT} =  \log \left(\sum_{m=0}^{M_I}\sum_{m'=0}^m e^{\left( \frac{\Delta G_m(f) - \Delta G_{m'}(f)}{k_BT}\right)} \right)
\end{equation}
\begin{equation}\label{Beff_TheoIU}
    \frac{B_{IU}(f)}{k_BT} =  \log \left(\sum_{m=M_I+1}^{M}\sum_{m'=M_I+1}^m e^{\left( \frac{\Delta G_m(f) - \Delta G_{m'}(f)}{k_BT}\right)} \right)
\end{equation}
\end{subequations}
where the double sum runs over all hairpin configurations, labeled by $m$ and $m'$, being $M_I$ the number of unzipped base-pairs (bp) at $I$ and $M$ the total number of bp.

The resulting barriers are shown in Fig. \ref{fig2}d (solid symbol: pulling; empty symbol: hopping), while the extracted values for $\Delta G^0_{ij}$ and $k_0^{ij}$ are summarized in Table \ref{Tab_all}. We find good agreement with theoretical predictions.

\begin{figure*}
    \centering
    \includegraphics[]{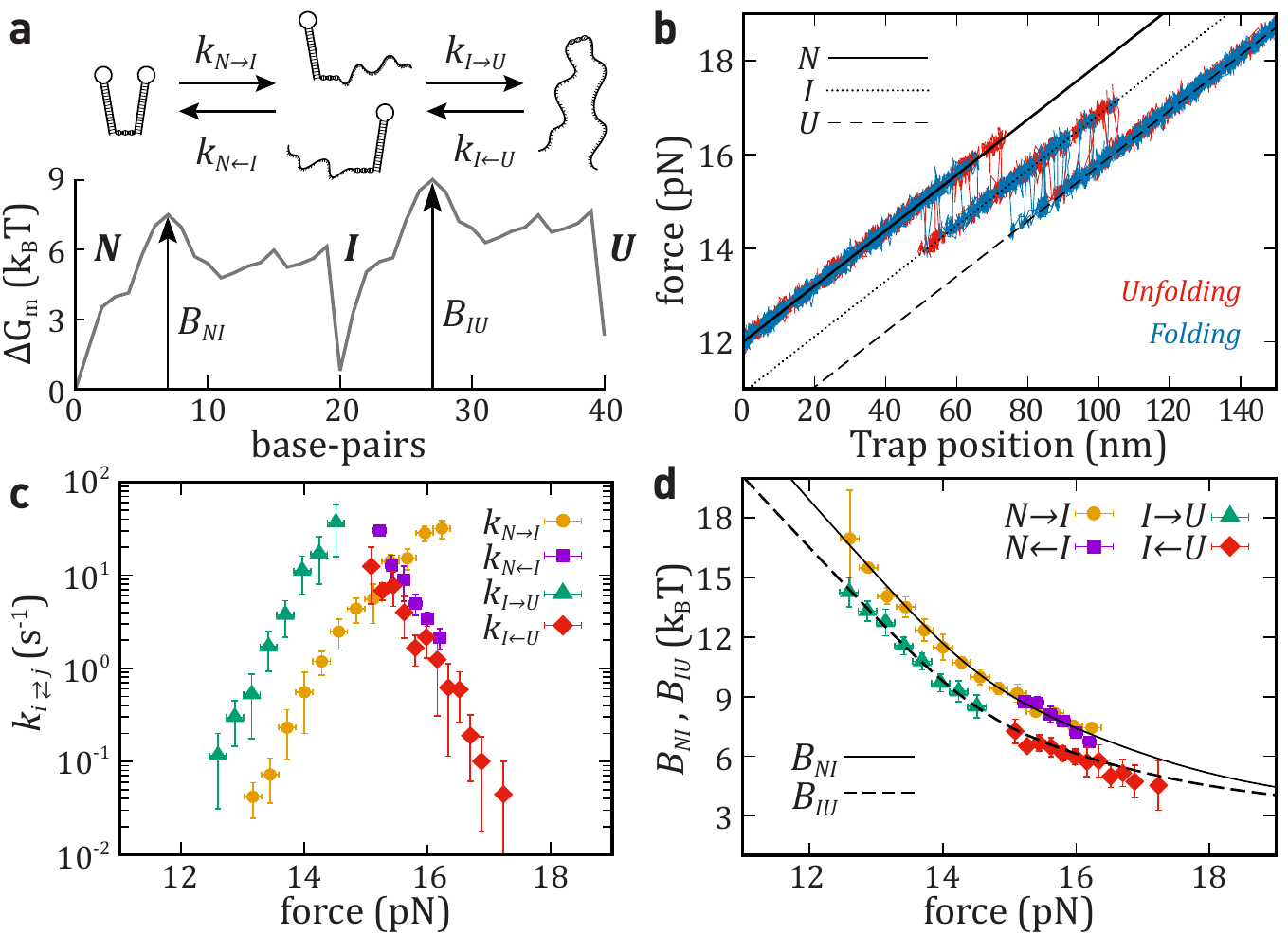}
    \caption{\textbf{Unfolding-folding kinetics of a doubly degenerate intermediate (hairpin HI2)} (a) Top: Schematic unfolding and folding pathways for HI2 that has two degenerate intermediates (both hairpins have the same sequence) depending which of the two hairpins is unfolded. Bottom: mFEL ($\Delta G_m$) as a function of the number of unfolded base-pairs ($m$) at $15$pN. The barriers, $B_{NI}$ and $B_{IU}$, are highlighted (black arrows). (b) Five unfolding (red) and folding (blue) FDCs (pulling speed equals $100$nm/s,  each trajectory taking $\sim$ 2s.). Force branches for states $N$ (black solid line), $U$ (black dashed line) and $I$ (dotted line). (c) Kinetic rates of unfolding: $k_{N \to I}$ (yellow circle) and $k_{I \to U}$ (green triangle); and folding: $k_{N \leftarrow I}$ (purple square) and $k_{I \leftarrow U}$ (red diamond). (d) Barriers mediating transitions between $N$ and $I$ (black solid line) and between $I$ and $U$ (black dashed line) predicted from Eqs.(\ref{Beff_TheoNI},\ref{Beff_TheoIU}) and compared with the experimental results (symbols). The results shown in panels (c) and (d) are the average over four different molecules and the error bars correspond to the statistical errors.}
    \label{fig4}
\end{figure*}

%
{\bf A doubly degenerate intermediate and two folding pathways.} Next, we designed hairpin HI2, which contains two identical DNA hairpins serially connected and separated by a short (29bp) double-stranded DNA segment (Fig. \ref{fig1}c). The native hairpin $N$ can unfold via two different pathways, each characterized by an intermediate corresponding to the unfolding of one of the two hairpins (Fig. \ref{fig4}a). However, as both hairpins are identical, they cannot be experimentally distinguished. Therefore, we define a global intermediate $I$ comprising the two intermediates.

The mFEL of HI2 is defined as the potential of mean force where a given number $m$ of open bps ($0\le m\le 40$) is distributed among the two hairpins. The mFEL shows a single intermediate at $m = 20$ (Fig.~\ref{fig4}a-bottom), where one hairpin is folded, and the other is unfolded.

In Fig.~\ref{fig4}b we show unfolding (red) and folding (blue) FDCs. Like for HI1 there are three force branches for states $N$, $I$, and $U$ (black lines). We use eCEBA to determine the force-dependent kinetic rates (Fig.~\ref{fig4}c) and the effective barriers $B_{NI}$ and $B_{IU}$ mediating transitions between the three states (Fig.~\ref{fig4}d). Results for the folding free energies and attempt rates are shown in Table \ref{Tab_all} (middle). Note that, although both hairpins are identical, the barriers for $N\rightleftarrows I$ and $I\rightleftarrows U$ are different (Fig.~\ref{fig3}d).

\begin{figure*}
    \centering
    \includegraphics{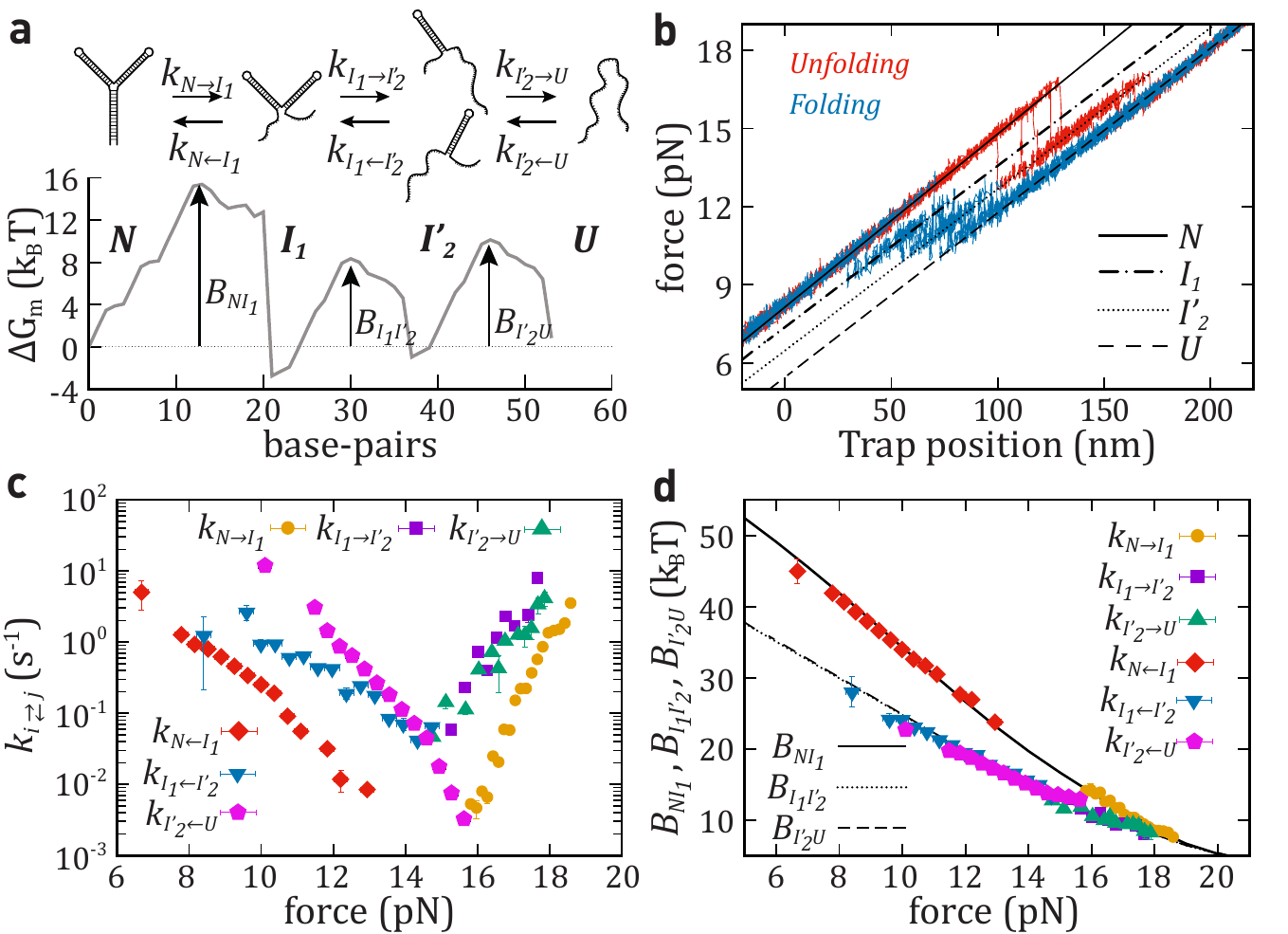}
    \caption{\textbf{Unfolding-folding kinetics of a triple intermediate (hairpin HI3).} (a) Top: Schematic unfolding and folding pathways for HI3. Bottom: mFEL ($\Delta G_m$) as a function of the number of unfolded base-pairs ($m$) at $15$pN. The barriers, $B_{NI_1}$,  $B_{I_1,I'_2}$ and $B_{I'_2U}$, are highlighted (black arrows). (b) Five unfolding (red) and folding (blue) FDCs (pulling speed equals $100$nm/s, each trajectory taking $\sim$ 2s.). Force branches for states $N$ (black solid line), $U$ (black dashed line), $I_1$ (dotted-line line) and $I_2$ (dotted line). (c) Kinetic rates of unfolding: $k_{N \to I_1}$ (yellow circle), $k_{I_1 \to I'_2}$ (purple square), and $k_{I'_2 \to U}$ (green triangle); and folding: $k_{N \leftarrow I_1}$ (red diamond), $k_{I_1 \leftarrow I'_2}$ (blue down-pointing triangle), and $k_{I'_2 \leftarrow U}$ (pink pentagon). (d) Barriers mediating transitions between $N$ and $I_1$ (black solid line), between $I_1$ and $I'_2$ (black dotted line), and between $I'_2$ and $U$ (black dashed line) predicted from Eqs.(\ref{Beff_TheoNI},\ref{Beff_TheoIU}) extended to one more intermediate and compared with the experimental results (symbols). The results shown in panels (c) and (d) are the average over five different molecules and the error bars correspond to the statistical errors.}
    \label{fig5}
\end{figure*}
%
{\bf Three intermediates and two folding pathways.} The last studied molecule (HI3) is a DNA three-way junction  (Fig.~\ref{fig5}a). For the first intermediate, $I_1$, the 20 bp of the main stem (before the junction) are unzipped. Further unzipping of HI3 distributes open bps between the two upper arms of HI3. The calculated mFEL (Fig.~\ref{fig5}a, bottom) shows two additional intermediates, each for the unfolding of one arm. Therefore, HI3 can take two different pathways to unfold starting from $I_1$: $I_1 \to I_2 \to U$ or $I_1 \to I_3 \to U$ depending on which arm is opened first. Since we cannot distinguish between $I_2$ and $I_3$ from the FDCs (Fig.~\ref{fig5}b), we studied the unfolding and folding pathway as $N \leftrightarrows I_1 \leftrightarrows I'_2 \leftrightarrows U$. Here, $I'_2$ comprises $I_2$ and $I_3$: $I'_2=I_2 \cup I_3$. 

In Fig.~\ref{fig5}c we show the six kinetic rates of HI3. From Eq.~(\ref{Beff_Unf_exp},~\ref{Beff_Ref_exp}), we derive the effective barrier, the free energy difference, and attempt rates for $N\leftrightarrows I_1$, $I_1\leftrightarrows I'_2$, and $I'_2\leftrightarrows U$ (Table \ref{Tab_all}). In Fig.~\ref{fig5}d we show $B_{N I_1}(f)$, $B_{I_1 I'_2}(f)$, and $B_{I'_2 U}(f)$ together with the theoretical prediction from Eqs.~(\ref{Beff_TheoNI},~\ref{Beff_TheoIU}) extended to include a second intermediate. Due to the sequence similarity between the two arms in the three-way junction, the barriers for $I'_2\leftrightarrows U$ and $I_1\leftrightarrows I'_2$ are nearly equal.

%
{\bf Discussion.} In the present work, we used non-equilibrium pulling experiments to determine the force-dependent unfolding/folding kinetic rates for DNA hairpins with three different kinds of intermediates (Fig.~\ref{fig1}c): a hairpin with an inner-loop and a single intermediate (HI1); a two-hairpin structure with a doubly degenerate intermediate (HI2); and a three-way junction with three intermediates (HI3). For hairpin HI1, we also derived the kinetic rates from equilibrium hopping experiments. We showed that pulling experiments recover kinetic rates at forces where intermediates cannot be sampled in equilibrium conditions. In general, the force gap between the unfolding and folding forces facilitates reconstructing the kinetic barrier $B_{ij}(f)$ in a larger force range \cite{alemany2016}. Further extension of the range of forces where kinetic rates are measured might be achieved by increasing (decreasing) the loading (unloading) rate during the unfolding (folding) process. The simplicity of the BE model \cite{bell1978, evans1997, evans2001} makes it a preferred model to fit the kinetic rates. Here we exploited eCEBA \cite{manosas2006, bizarro2012, alemany2016, alemany2017} to measure the force-dependent kinetic barriers, $B_{ij}$, and the free energy differences, $\Delta G_{ij}$, between different states. Our results showed good agreement between the experimental values and the predictions based on the nearest neighbor model (Table~\ref{Tab_all}). The folding free-energy values per bp of the nearest-neighbor model used in the comparison are obtained from the Mfold. The latter uses energy parameters derived from temperature melting data collected in calorimetry (bulk) experiments \cite{santalucia1998,zuker2003}. The good agreement between the measured force-dependent kinetic barriers and the KD model prediction allowed us to estimate values for attempt rates for native, intermediate and unfolded states. Attempt rates are important for molecular dynamic simulations, where timescales need to be set properly. In general, the KD model has more predictive power than the BE model which assumes a single kinetic barrier between states. In contrast, in the KD model, folding is a diffusive process in a one-dimensional mFEL with many intermediate configurations. In principle, Eq.\eqref{Beff_TheoNU} (two-states) and Eq. \eqref{Beff_TheoNI},\eqref{Beff_TheoIU} (three states) might be inverted (by discretizing the force range) to derive the energy set, $\Delta G_m^0$, directly from the measured $B_{ij}(f)$.

\begin{table}[ht]
\begin{tabular}{ccccc}
                     &              & $k_0^{ij}$ ($s^{-1}$)  & \multicolumn{2}{c}{$\Delta G_{i,j}^0$ ($k_BT$)} \\ \cline{3-5} 
                     & $i, j$        &                         & Exp.             & Pred.                 \\ \hline \hline
\multirow{4}{*}{HI1}  & $N$, $I$      & $(5 \pm 1) \times 10^7$ & $30 \pm 2$             & $30 \pm 1$             \\
                & $I$, $U$      & $(7 \pm 1) \times 10^6$ & $27 \pm 3$             & $28 \pm 1$            \\ \cline{2-5}
 & $N$, $I$      & $(5 \pm 1) \times 10^7$ & $31 \pm 3$             & $30 \pm 1$             \\
                   & $I$, $U$      & $(6 \pm 1) \times 10^6$ & $28 \pm 4$             & $28 \pm 1$             \\ \hline
\multirow{2}{*}{HI2} & $N$, $I$      & $(5 \pm 1) \times 10^5$ & $54 \pm 2$             & $52 \pm 2$             \\
                     & $I$, $U$      & $(2 \pm 1) \times 10^6$ & $51 \pm 1$             & $55 \pm 2$             \\ \hline
\multirow{3}{*}{HI3}&  $N$, $I_1$    & $(6 \pm 1) \times 10^5$ & $57 \pm 4$             & $52 \pm 2$             \\
                 & $I_1$, $I'_2$ & $(2 \pm 1) \times 10^6$ & $38 \pm 3$             & $41 \pm 2$             \\
                    & $I'_2$, $U$   & $(9 \pm 2) \times 10^5$ & $39 \pm 2$             & $40 \pm 2$            
\end{tabular}
\caption{Folding free energies and kinetic attempt rates for the three studied hairpins. The results of molecule HI1 in top (bottom) rows correspond to pulling (hopping) experiments. The error bars for the experimental values correspond to the statistical error considering all studied molecules, while the error bar in the Mfold prediction corresponds to the standard error considering several experimental values.}\label{Tab_all}
\end{table}

Notice that HI2 and HI3 were designed to have degenerated and indistinguishable folding intermediates. For non-degenerate and distinguishable intermediates, the analysis of the respective folding pathways follows the same steps as we did for HI1  (Fig. \ref{fig2}). However, for non-degenerate and indistinguishable intermediates, dynamics might not be Markovian and the KD model (c.f. Eq.\eqref{Beff_TheoNU}) should be revisited.

In cases where the mFEL is not known (e.g., in tertiary RNAs and proteins), Eq.\eqref{Beff_TheoNU} is inapplicable. However, one can still reconstruct the kinetic barrier by using Eq.\eqref{k_Unf} from the measured $k_{N\to U}(f)$ and the knowledge of $k_0$. The latter can be obtained from the extrapolated value of $B_{NU}(f)$ to zero force which is approximately equal to $\Delta G_{NU}^0$, $k_0\simeq k_{N\to U}(f=0)\exp(\frac{\Delta G_{NU}^0}{k_BT})$. Determining $k_{N\to U}(f=0)$ requires reconstructing the kinetic rates at low forces. In section S2 of Supp. Info., we describe the procedure used to reconstruct the kinetic rates and barriers down to zero force. Results are tested for hairpin HI1 finding results in agreement with those summarized in Tab. \ref{Tab_all}.

Future studies might address kinetic barrier measurements  at different temperatures\cite{mao2005, stephenson2014, sara2015, rico2018} to separate the enthalpic and entropic contributions. These studies might be applied to other non-native states, e.g., misfolded structures. eCEBA might also find applications to unravel the kinetic role of complex molecules, such as chaperons and other enzymes that facilitate molecular folding/unfolding reactions, and ligand binding. The possibility of characterizing changes in the kinetic barrier's height as a function of force under different conditions (e.g., crowding, binding agents, temperature, ionic strength) will permit us to understand how molecular machines in cells respond to external signals and perturbations. 

\section*{Supporting Information}
Additional theoretical details, and methods are provided in the Supporting Information document. In section S1, it is presented a detailed derivation of the kinetic barrier in the KD model. In section S2, it is shown how to reconstruct the kinetic rates for complex molecules.

\section*{Acknowledgments}
M.R. and F.R. acknowledge support from European Union’s Horizon 2020 Grant No. 687089, Spanish Research Council Grants FIS2016-80458-P, PID2019-111148GB-I00 and ICREA Academia Prizes 2013 and 2018.\\

\bibliography{biblio}

\end{document}


\maketitle

\section{Derivation of the kinetic barrier in the KD model}
\begin{figure}[h]
    \centering
    \includegraphics{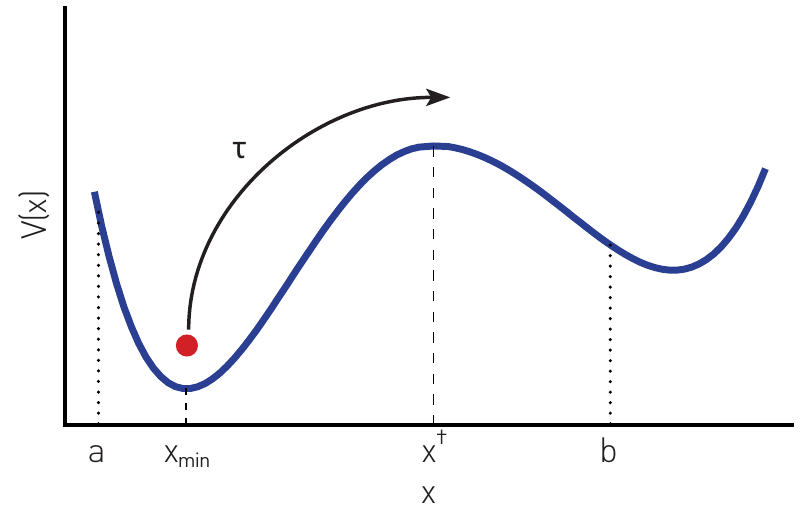}
    \caption{\textbf{Brownian particle in a double well potential.} The minim and maximum values of $V(x)$ are located at $x_{min}$ and $x^\dagger$, respectively. Positions $a$ and $b$ are used to model the absorbing and reflecting boundaries of the dynamics of the particle (red dot). }
    \label{fig6}
\end{figure}

Let us consider the system presented in Fig. \ref{fig6}, where a Brownian particle diffuses in one dimensional potential of mean force $V(x)$. 
The time evolution of the probability density function $p(x,t)$ to find the particle at position $x$ at time $t$ follows the Fokker-Planck equation\cite{zwanzig2001,hyeon2007b},
\begin{equation}\label{FP_eq}
    \frac{\partial p(x,t)}{\partial t} = D \frac{\partial}{\partial x}\left[ \frac{\partial}{\partial x} + \frac{1}{k_BT} \frac{dV(x)}{dx} \right]p(x,t)
\end{equation}
where $D= k_BT/\gamma$ is the diffusion coefficient, $\gamma$ is the friction coefficient, $k_B$ is the Boltzmann constant, and $T$ is the temperature.

By considering that the particle at time $t=0$ is located in a region $\mathcal{R}=[a,b]$ at position $x_0\in \mathcal{R}$, then $p(x,0) = \delta(x-x_0)$. Considering the survival probability defined as the probability of the Brownian particle to remain inside $\mathcal{R}$ at time $t$, $\mathcal{S}_{x_0}(t,\mathcal{R}) = \int_{x\in \mathcal{R}} p(x,t)dx$, it is possible to determine the density function of the survival time, which is equal to $-\frac{\partial \mathcal{S}_{x_0}(t,\mathcal{R})}{\partial t}$. The mean first passage time $\tau$ is defined as $\tau(x_0) = -\int_0^\infty dt \ t \ \frac{\partial \mathcal{S}_{x_0}(t,\mathcal{R})}{\partial t}$. A differential equation can be written and solved for $\tau(x)$,  with absorbing ($\tau(b) = 0$) and reflecting ($\frac{\partial \tau}{\partial x}|_{x=a} = 0$) boundary conditions. The mean first passage time for a given position $x$ equals,
\begin{equation}
\tau(x) = \frac{1}{D}\int_x^b dy \ e^{\frac{V(y)}{k_BT}} \int_a^y dz \ e^{\frac{-V(z)}{k_BT}} \ .
\end{equation}
Finally, the kinetic rate to unfold $k_\to$ is defined as the inverse of the mean first passage time evaluated at $a=x_{min}$ ($k_\to = \tau(a)^{-1}$),
\begin{equation}\label{kU_Kram}
k_\to = \frac{D}{\int_a^b dy \ e^{\frac{V(y)}{k_BT}} \int_a^y dz \ e^{\frac{-V(z)}{k_BT}}} \ .
\end{equation}
By equating Eq. \eqref{kU_Kram} and Eq. (1a) we get:
\begin{equation}\label{Bij_Kram}
B = k_BT \left[ \log \left(\frac{k_0}{D} \int_a^b dy \ e^{\frac{V(y)}{k_BT}} \int_a^y dz \ e^{\frac{-V(z)}{k_BT}}  \right) \right] .
\end{equation}
For a DNA/RNA molecule forming a hairpin structure, the reaction coordinate $x$ for the unfolding-folding reaction (N$\leftrightarrow$U) equals the number of released base-pairs \textit{m}. As \textit{m} is an integer, the double integral in Eq. \eqref{Bij_Kram} becomes a double sum with limits $a \equiv 0$ and $b \equiv M$. These limits correspond to the configurations where the molecule is folded in N ($m=0$) and unfolded in U ($m=M$), respectively. Intermediate \textit{m} values ($0<m<M$) indicate partially folded configurations. Taking $dy,dz\equiv l_0$ equal to the inter-monomer distance, we can rewrite Eq. \eqref{Bij_Kram} as,
\begin{equation}\label{BNU_Kram}
B = k_BT \left[ \log \left(\frac{k_0l_0^2}{D} \sum_{m=0}^M \sum_{m'=0}^{m} e^{\frac{V_{m} - V_{m'}}{k_BT}}  \right) \right]
\end{equation}
The diffusion coefficient satisfies $D\simeq k_0l_0^2$, therefore we get Eq. (4). \\


\section{Reconstruction of the kinetic rates}
To determine $k_0$, we use the fact that the kinetic barrier to unfold at zero force can be approximated by the folding free energy, i.e., $B_{ij}(f=0) \simeq \Delta G_{ij}^0$. This result can be obtained from Eq. (4) by calculating $B_{ij}(0)$ for $\Delta G_m^0=mg$ with $g>0$ the average free energy per bp.  Extrapolating  $k_{i\to j}(f)$ to zero force, we can determine the attempt rate $k_0$ using Eq. (1a), 
\begin{equation} \label{k_Unf2}   
k_0\simeq k_{i\to j}(f=0)\exp{\Bigl(\frac{\Delta G_{ij}^0}{k_BT}\Bigr)} \ .
\end{equation}
To extrapolate the $k_{i\to j}(f)$ to zero force, we need to reconstruct the kinetic rates at sufficiently low forces. A possible strategy is to take advantage of the detailed balance condition by merging Eqs. (1b) and (2),
\begin{equation}\label{eq:DBC2}
    \log \left( \frac{k_{i \leftarrow j}(f)}{k_{i\to j}(f)} \right) =    \frac{1}{k_BT} \left[ \Delta G_{ij}^0 - \int_0^f (x_j(f')-x_i(f'))df' \right]\ ,
\end{equation}
with $x_i$ ($x_j$) being the extension of the initial (final) state, $\Delta G^0_{ij}$ the folding free-energy, and $k_{i\to j}(f)$ ($k_{i \leftarrow j}(f)$) the force-dependent unfolding (folding) kinetic rate.

\begin{figure}[h]
    \centering
    \includegraphics{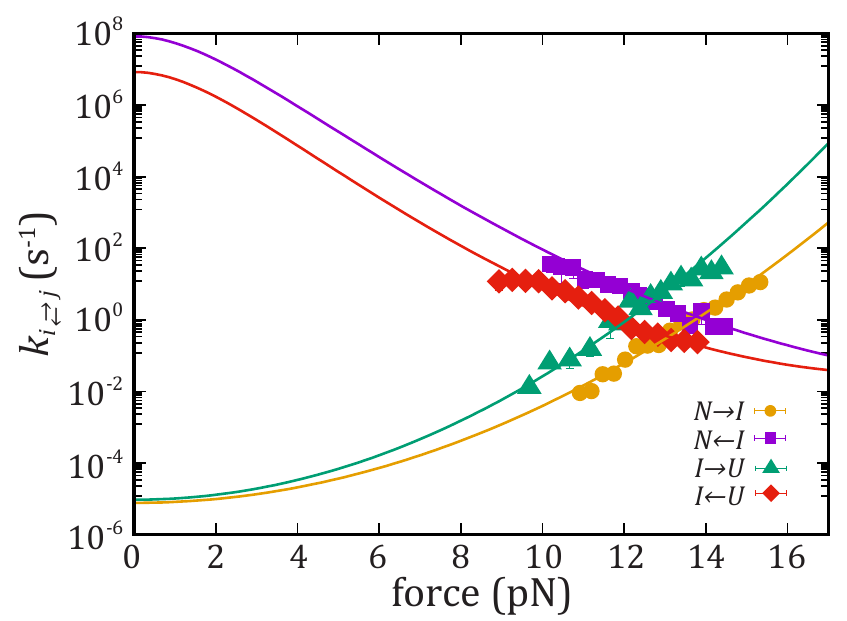}
    \caption{\textbf{Reconstructed kinetic rates.} Symbols correspond to the experimental values and solid lines correspond to the reconstructed kinetic rates.}
    \label{fig7}
\end{figure}

To illustrate and test the approach, we have analyzed the HI1 molecule by extrapolating the unfolding and folding kinetic rates to zero force. The extrapolation has been done by fitting simultaneously the unfolding kinetic rate to a quadratic function (in $\log k$ versus $f$ scale) and imposing Eq. \eqref{eq:DBC2} to reconstruct the folding rate. The term, $\int_0^f (x_j(f')-x_i(f'))df'$ in Eq. \eqref{eq:DBC2} is determined by using the elastic parameters of ssDNA \cite{ViaderGodoy2021}. The extrapolated unfolding and folding kinetic rates of HI1 are shown as solid lines in Fig. \ref{fig7} for $i,j= \{N, \ I, \ U\}$. Let us mention that the quadratic fits to the unfolding rates in a log-normal plot agree with the curvature reported in a previous study using a similar approach \cite{Rognoni2012}.
The extrapolated kinetic rate values from $k_{N\to I}(f)$ (yellow line in Fig. \ref{fig7}) and $k_{I\to N}(f)$ (green line in Fig. \ref{fig7}) are equal to $k_{N\to I}(0)=(7.7\pm 1)\times 10^{-6} s^{-1}$ and $k_{I\to N}(0)=(9.5\pm 1)\times 10^{-6} s^{-1}$. Using the average experimental folding free energy values reported in Tab. (1) and Eq. \eqref{k_Unf2}, we get $k_0^{NI}=(8\pm 1)\times10^{7} s^{-1}$ and $k_0^{IU}=(8\pm 1)\times10^{6} s^{-1}$, which agree pretty well with the values summarized in Tab. (1) obtained by matching the experimental data to the predicted kinetic barriers (Eqs. (9a) and (9b)).

\bibliography{biblio}